\documentclass{optica-article}

\journal{opticajournal} 

\articletype{Research Article}

\usepackage{lineno}

\begin{document}

\title{A Residual-Subspace Constraint Framework for Fourier Ptychographic Microscopy}

\author{Sui-peng Wang\authormark{1}, Si-yi Xie\authormark{1}, Chang-tao Cai\authormark{1}, Zhun Wei\authormark{2}, and Rui Chen\authormark{1,*}}

\address{
\authormark{1}{School of Physics \& State Key Laboratory of Optoelectronic Materials and Technologies, Sun Yat-sen University, Guangzhou, 510275, China}\\
\authormark{2}{College of Information Science and Electronic Engineering, Zhejiang University, Hangzhou 310027, China}
}
\email{\authormark{*}chenr229@mail.sysu.edu.cn} 


\begin{abstract*} 
The reconstruction fidelity of computational optical imaging is fundamentally constrained by the model-reality gap, i.e., the inevitable discrepancy between idealized forward models and the physical imaging process. Conventional paradigms attempt to bridge this gap through exhaustive system calibration or explicit parameter estimation, which are often computationally intensive and prone to severe non-convex stagnation. This paper introduces a Residual-Subspace Constraint Framework (RSCF) to achieve robust Fourier ptychographic microscopy. Instead of treating residuals as unstructured errors, RSCF leverages subspace decomposition to decouple low-rank, systematic mismatches from stochastic noise, thereby isolating stable information manifolds that remain invariant to forward-model inaccuracies. By embedding this subspace constraint into the iterative engine, the framework selectively suppresses error-amplifying components, enabling high-fidelity phase and amplitude recovery without explicit hardware calibration. Numerical simulations and experimental validations demonstrate that RSCF yields superior convergence acceleration and artifact suppression under severe optical aberrations and LED misalignment. This information-centric paradigm provides a versatile, model-agnostic strategy to enhance robustness across diverse computational imaging modalities.
\end{abstract*}

\section{Introduction}
Computational optical imaging (COI) combines physical imaging formation with numerical reconstruction to overcome the performance of conventional optical systems \cite{Mait:18, Liu2024_AI_1_012001}. Through the joint design of optical acquisition and computational inversion \cite{Bian2025_LSA_14_162, Wang2025_Optica_12_113}, COI technologies have enabled imaging performance beyond the limits of conventional hardware architectures, including super-resolution imaging \cite{tian2011survey}, large space-bandwidth product \cite{Zheng:13}, extended depth of field \cite{10.1145/3197517.3201333}, quantitative phase recovery \cite{doi:10.1021/acsnano.1c11507} and low-light imaging \cite{chen2018learning}. Despite the diversity of imaging modalities, most COI systems can ultimately be formulated as inverse problems that recover latent object information from physically encoded measurements \cite{Review_Zuo:2022}. Consequently, reconstruction quality is intrinsically governed by the accuracy of the assumed forward model and the stability of the associated optimization process \cite{Pan:19, Review_Zuo:2022}.

In practical scenarios, however, the assumed forward model inevitably deviates from the actual physical imaging process \cite{doi:10.1137/17M1141965}. Even under carefully calibrated conditions, reconstruction remains affected by optical aberrations, illumination fluctuations, calibration drift, coherence mismatch, and other hardware nonidealities \cite{chen2023differentiable}. These perturbations are further coupled with the intrinsic non-convexity and ill-posedness of the iterative inverse reconstruction \cite{sui2024non}, often leading to degraded convergence stability, artifact amplification, and degradation of reconstruction fidelity. As COI systems continue to evolve toward higher resolution, larger field of view, and increasingly complex acquisition strategies, robustness against forward-model mismatch has become a central challenge in practical COI \cite{doi:10.34133/adi.0117, 11445767}.

To address this challenge, extensive efforts have targeted explicit error modeling \cite{Chen:25,Chen:22} and regularized optimization \cite{zheng:21,Lee2024_OE_32_25343,Shi:21}, including statistical noise models \cite{Ma:24}, aberration correction \cite{oh2025digital}, calibration refinement \cite{daiToleranceawareDeepOptics2025}, adaptive optimization schemes \cite{Ou:22,9477112}, handcrafted priors \cite{6737048}, or deep learning-based reconstruction framework \cite{He:26,Barbastathis:19, CHEN2025112016}. While effective under specific assumptions, these methods generally rely on explicitly parameterizing individual error sources or enforcing rigid, predetermined priors. In realistic imaging scenarios, however, different error sources are frequently intertwined and dynamically coupled, making accurate analytical modeling difficult and limiting generalization under mixed-degradation conditions \cite{Pan2017_JBO_22_096005}. More recent alternatives attempt to bypass explicit calibration by performing optimization within the transformed feature domains \cite{Zhang:24,wu_waveletforward_2025,zhang_highfidelity_2025}. While empirically robust, these methods lack an explicit mechanism to characterize how modal mismatch propagates across different information channels. 

Conventional reconstruction frameworks routinely treat data residuals as pure stochastic noise, minimizing them indiscriminately through pixel-wise loss functions \cite{Review_Zuo:2022}. However, the residuals generated during iterative reconstruction are rarely completely random. Because they embody the combined effects of model mismatch, system-dependent perturbations, and reconstruction ambiguity, their distributions may exhibit distinct correlations and latent structural organization during iterative inversion. Elucidating whether this residual structure contains exploitable information represents an open and important challenge for developing robust computational imaging paradigms that transcend conventional noise-suppression limits.

In this work, we reveal that data residuals in iterative COI reconstruction exhibit a pronounced, low-rank spectral structure in the residual domain, rather than behaving as fully uncorrelated random noise. By analyzing the residual matrix constructed across iterative steps, we observe that systematic, model-mismatch energy tends to concentrate within a few dominant singular modes, whereas trailing low-energy components exhibit substantially weaker correlations and high stochasticity. Crucially, this spectral organization emerges directly from the residual statistic without requiring explicit modeling of individual error sources. Leveraging this low-rank property, we introduce a residual-subspace-constrained reconstruction framework for COI inverse problems. The architecture bypasses explicit error modeling entirely. Instead, it operates directly on the spectral representation of data residuals, embedding a principal component constraint into the optimization loop. Singular-spectrum analysis drives this process. By prioritizing the dominant residual modes, which encode structured mismatch and object ambiguity, the proposed framework suppresses highly stochastic, low-energy components. This targeted filtering guides the optimization path toward true physical object features, leaving systematic model inconsistencies safely confined within the loss space.

To validate the proposed framework, we employ Fourier ptychographic microscopy (FPM) as a representative COI platform \cite{Zheng:13}. FPM serves as an ideal testbed for investigating residual-domain behaviors, as it couples synthetic aperture imaging with nonlinear phase retrieval through highly redundant intensity measurements \cite{xu2024fourier}. Its iterative reconstruction is highly sensitive to coupled model mismatches, including LED misalignment, optical aberrations, coherence deviations, and intensity fluctuations \cite{Sun:16,Ou:14}, all of which are easily compounded by the non-convex optimization dynamics. Simulation and experimental results validate the efficacy of the proposed framework. Under mixed degradation conditions, the architecture successfully suppresses artifact accumulation and noise amplification while faithfully preserving fine structural details. Beyond these empirical gains, this work redefines the role of data residuals in inverse problems. Rather than treating residuals as stochastic noise to be minimized blindly, we demonstrate that they serve as structured indicators of forward-model inconsistency. Capitalizing on this spectral organization offers a generalizable pathway to counter imperfect physical modeling, extending naturally to other COI technologies such as holography \cite{holography} and tomography \cite{tomography}.

\section{Methods}
\subsection{Forward model of FPM and Conventional reconstruction framework}
FPM reconstructs a wide-field, high-resolution complex field by sequentially illuminating the specimen with plane waves of varying incident angles. The modulated wavefront propagates through a low numerical-aperture (NA) objective and is recorded by an image sensor. Because conventional optical sensors record intensity rather than phase, the system captures only a sequence of low-resolution intensity measurements, denoted as $\mathbf{I}_n$. The forward physical imaging process for the $n$-th illumination angle is formulated as:
\begin{equation}
    \mathbf{I}_n = \left| \mathbf{F}^{-1} \mathbf{P} \mathbf{M}_n \mathbf{F} \mathbf{U} \right|^2 + \eta_n,
    \label{I_n}
\end{equation}
where $\mathbf{F}$ and $\mathbf{F}^{-1}$ denote the two-dimensional Fourier and inverse Fourier operators, respectively. $\mathbf{U}$ signifies the latent complex amplitude of the specimen and $\mathbf{P}$ denotes pupil function of the imaging system. $\mathbf{M}_n$ is the frequency-selection matrix corresponding to the shift induced by the $n$-th LED. $\eta_n$ models additive measurement noise.

The objective of FPM reconstruction is to retrieve the object's complex amplitude function $\mathbf{U}$ by solving an ill-posed inverse problem. From an optimization viewpoint, this phase-retrieval task is typically cast as minimizing discrepancy between the measured intensities and those predicted by the forward physical model \cite{Yeh:15}. In conventional FPM frameworks, the loss function is constructed directly in the spatial pixel domain based on amplitude or intensity residuals:
\begin{equation}
    \mathcal{L}_{\text{conventional}} (\mathbf{U},\mathbf{P}) = \left\| \mathbf{I}_n^{\gamma} - \left| \mathbf{F}^{-1} \mathbf{P} \mathbf{M}_n \mathbf{F} \mathbf{U} \right|^{2\gamma} \right\|_2^2,
    \label{eq:con_loss}
\end{equation}
where $\gamma$ acts as an intensity correction factor (typically $\gamma = 1$ for intensity-based or $\gamma = 0.5$ for amplitude-based optimization). Although regularization strategies can partially improve robustness, the underlying forward model fundamentally relies on the ideal assumptions of linearity and space-invariance \cite{Goodman:17}. In practice, these assumptions are often violated by uncalibrated aberrations, vignetting, partial coherence mismatch, and inherent system uncertainties. As a result, the residuals between measurements and model predictions contain not only stochastic noise, but also structured discrepancies induced by forward-model mismatch. Conventional pixel-wise losses treat these components equally, which can limit reconstruction fidelity and degrade convergence stability.

\subsection{Residual-domain statistical characterization}
Conventional FPM optimization evaluates reconstruction errors (Eq.\eqref{eq:con_loss}) directly in the spatial domain. However, the statistical properties of residuals differ substantially from those of raw images. Instead of being dominated by low-frequency redundancy, residual fields primarily contain high-frequency structured perturbations associated with model mismatch, optical aberration, structural distortion, and measurement uncertainty.

\begin{figure*}[thb!]
    \centering\includegraphics[width= \linewidth]{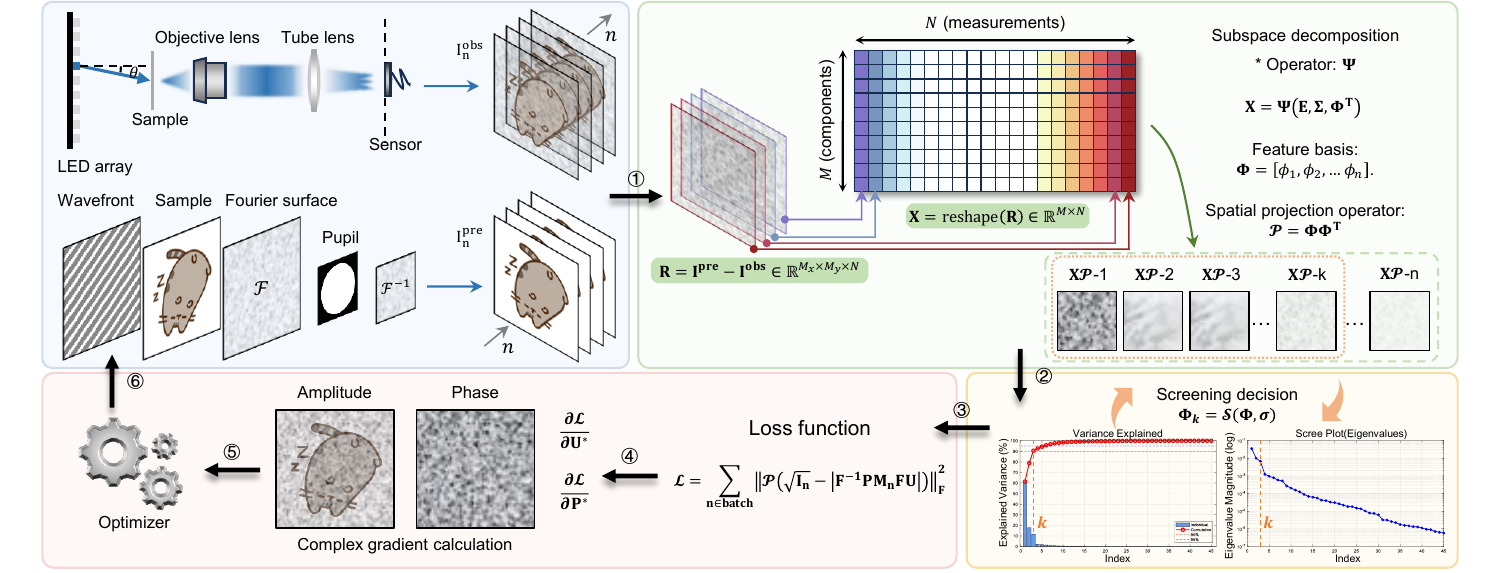}
    \caption{RSCF reconstruction workflow. Consisting of six steps: 1) generate low-resolution images from current estimate and compute residual images by comparison with experimental measurements; 2) perform subspace decomposition on the residual matrix formed by the spatial reshaping operator and select appropriate filtering operators to partition the residual subspace; 3) compute the reconstruction error in the subspace; 4) evaluate the gradient of the optimization objective with respect to the complex amplitude by backpropagating the error; 5) apply an appropriate optimizer to update the gradients; and 6) update the system parameters to refine the predicted wavefront.
    }
    \label{figure 1}
\end{figure*}

To characterize these structures, as shown in Fig. \ref{figure 1}, we define the residual tensor between the predicted and measured intensities as:
\begin{equation}
    \mathbf{R}=\textbf{I}^\text{pre}-\textbf{I}^\text{obs} \in \mathbb{R}^{M_x \times M_y \times N}, 
    \label{eq:Residual}
\end{equation}
where $M_x$ and $ M_y$ denote the spatial dimensions of each low-resolution measurement. The residual tensor is reshaped into a matrix $\mathbf{X} \in \mathbb{R}^{M \times N}, M = M_x \times M_y$, such that each column corresponds to the spatial residual distribution under a specific illumination angle and each row delineates the evolution along the observation (illumination angle) dimension.

A subspace decomposition operator $\Psi$ is then applied to decompose the residual matrix $\mathbf{X}$ into a set of orthogonal basis vectors:
\begin{equation}
    \mathbf{X} = \Psi \left( \mathbf{E}, \mathbf{\Sigma}, \mathbf{\Phi}^\text{T} \right),
    \label{subspace decomposition}
\end{equation}
where $\mathbf{\Phi} = [\phi_1, \phi_2, \dots, \phi_N]$ constitutes the observation-dependent feature bases and $\Sigma$ contains the corresponding spectral responses. Dominant components are selected through a subspace-selection operator $\mathcal{S}$, 
\begin{equation}
\mathbf{\Phi}_k = \mathcal{S}(\mathbf{\Phi}, \sigma),
\label{S}
\end{equation}
where $\sigma$ represents a decision threshold derived from information-theoretic criteria, such as the cumulative variance contribution rate (CVCR) or the inflection point of the scree plot \cite{jolliffe2016principal}. Based on the truncated bases $\mathbf{\Phi}_k$, the orthogonal projection operator is given by:
\begin{equation}
\mathcal{P} = \mathbf{\Phi}_k \mathbf{\Phi}_k^\text{T}.
\label{Poperator}
\end{equation}

In this work, the decomposition operator $\Psi$ is implemented via principal component analysis based on singular value decomposition (SVD) \cite{doi:10.1137/1035134,doi:10.1137/090771806}, with comprehensive mathematical details and implementation provided in Section 1 of the Supplemental Document. To validate this approach, the statistical behavior of  residuals under isolated and hybrid systematic perturbations was investigated. Fundamentally, under an idea forward model, residuals should manifest as spatially uncorrelated fluctuations. However, persistent spatial correlations inevitably arise from experimental system imperfections or model mismatches. 
Because systematic errors exhibit strong correlations across both spatial domains and multi-angle measurements, their energy after matrix decomposition concentrates within a low-dimensional subspace, characterized by a few dominant high-energy components in the singular-value spectrum. Conversely, stochastic noise maintains a relatively uniform and flat spectral distribution. This distinctive spectral divergence, validated by the numerical results in Figs. S1 and S2 (Section 2 of the Supplemental Document), demonstrates that model-induced discrepancies are effectively confined to a low-dimensional residual subspace.

\subsection{Residual Subspace Constrained Framework for FPM}
Based on the residual subspace analysis, we introduce the Residual Subspace Constrained Framework (RSCF) for FPM. Rather than minimizing pixel-wise residuals directly in the spatial observation domain, the proposed RSCF-FPM evaluates and constrains reconstruction errors within the dominant residual subspaces:
\begin{equation}
    \mathcal{L}_{\text{RSCF}} (\mathbf{U},\mathbf{P}) = \sum_{n \in \text{batch}}\left\| \mathcal{P} \left( \sqrt{\mathbf{I}_n} - \left| \mathbf{F}^{-1} \mathbf{P} \mathbf{M}_n \mathbf{F} \mathbf{U} \right| \right)\right\|_\text{F}^2,
    \label{PCA}
\end{equation}
where $||\cdot||_\text{F}$ denotes the Frobenius norm,and $n$ indexes the measurements within a selected mini-batch. $\mathcal{P}$ denotes the projection operator constructed from the selected dominant subspace bases derived in Eq.\eqref{Poperator}.

By embedding this subspace projection directly into the optimization loop, the objective function shifts the evaluation of optimization residuals from the conventional pixel domain to a structured singular-value representation. This constraint selectively prioritizes the leading residual modes that encode structured model mismatch and object ambiguity, while systematically suppressing isotropic, low-energy stochastic noise. Consequently, the optimization trajectory is robustly guided toward true physical object features, preventing stochastic fluctuations from corrupting the reconstructed complex amplitude.

To prevent over-filtering and preserve fine, high-frequency structural details of the object, the decision threshold $\sigma$ is dynamically cross-referenced with the elbow point of the singular value spectrum. Furthermore, evaluating the projection operator over multiple randomly selected mini-batches and performing mini-batch gradient descent mitigates the risk of discarding salient, high-frequency spatial information. This stochastic optimization approach smoothens the error landscape, assisting the algorithm in bypassing poor local minima and improving global convergence stability \cite{Ruder16}.

Based on the loss function of Eq. \eqref{PCA}, the overall reconstruction workflow of the proposed RSCF-FPM is illustrated in Fig. \ref{figure 1}. The initial complex object estimate $\mathbf{U}_0$ is generated by upsampling the low-resolution intensity image captured under central LED illumination. Spatial residual images are computed iteratively by comparing the forward model predictions with the experimental data according to Eq. \eqref{eq:Residual}. 

During iterations, these residuals are randomly grouped into mini-batches, where the batch size controls the transition between stochastic and global gradient descent strategies. Subspace decomposition is performed within each batch, and the top $k$ spatial components selected by the threshold $\sigma$ are used to construct the projection domain loss in Eq. \eqref{PCA}. The resulting gradients are backpropagated through the optical forward model to simultaneously update the complex object function $\mathbf{U}$ and the pupil function $\mathbf{P}$ via the RMSprop optimizer. The detailed mathematical derivations for this optimization step are provided in Section 3 of the Supplemental Document. RMSprop adaptively scales the learning rate for non-stationary objective, effectively stabilizing gradient trajectories and accelerating convergence across complex optimization landscapes \cite{Liao:25}.

\section{Result} \label{sec:Result}
\subsection{FPM system Setup}
\begin{figure*}[thb!]
    \centering\includegraphics[width=\linewidth]{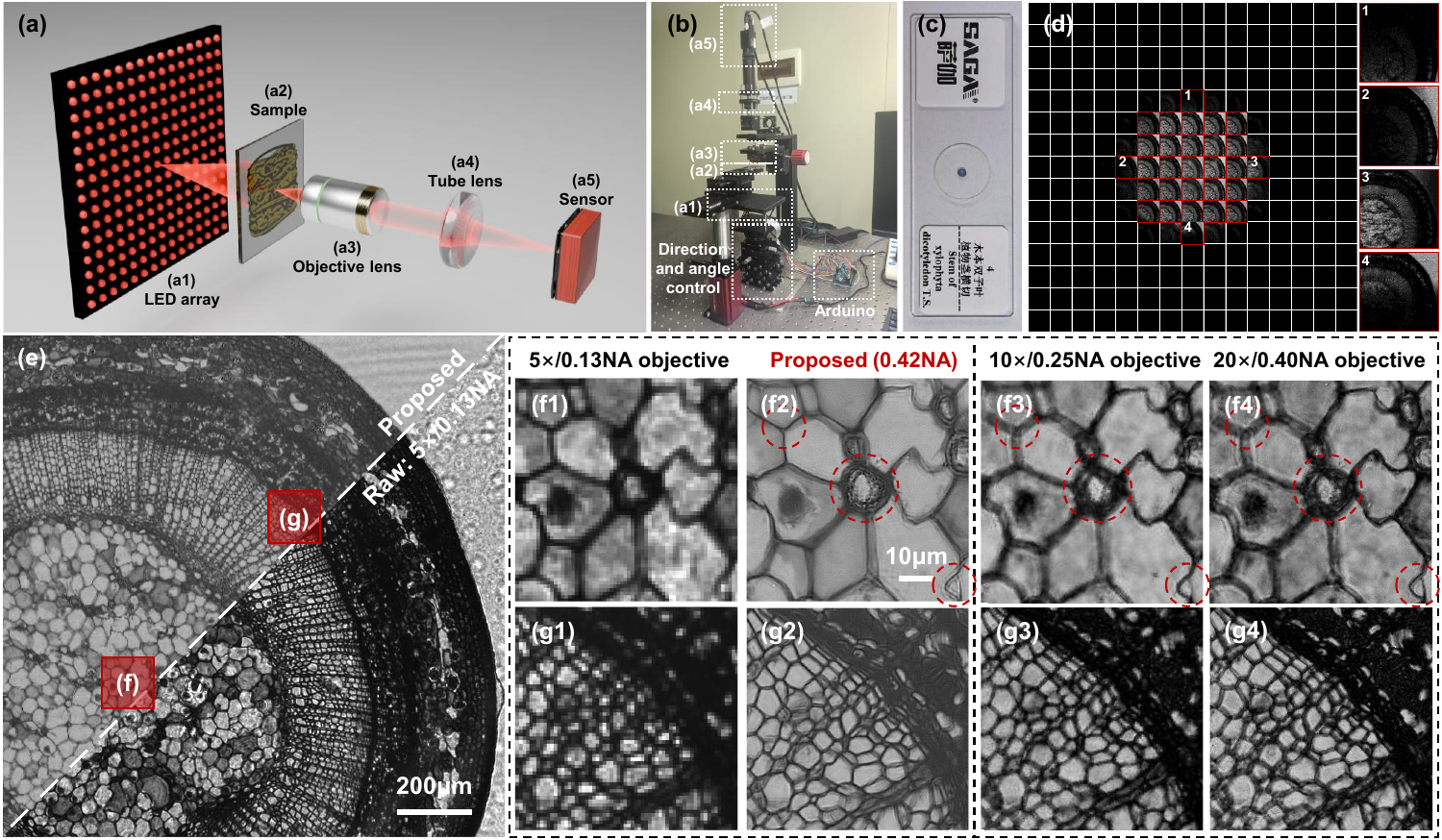}
    \caption{Full-FOV high resolution reconstruction of a plant stem cross-section. 
    (a) Three-dimensional schematic of the custom-built FPM system hardware. 
    (b) Physical implementation of the FPM platform, illustrating the synchronization network where an Arduino microcontroller delivers external TTL triggers to coordinate LED array illumination sequences and camera sensor exposures under automated host computer control. 
    (c) Stem of xylpophyta dicotyledon T.S. used as the experimental sample. 
    (d) A matrix of 225 sequential low-resolution images captured under varying red LED illumination angles, where the red-bounded matrix blocks identify bright-field frames exhibiting distinct vignetting boundaries. 
    (e) Comparison between a raw central-illumination image and the full-FOV high-contrast complex field image reconstructed using the proposed RSCF, with two structurally similar regions of interest (ROIs) designated as (f) and (g). 
    (f1)-(f4) Magnified views of ROI (f): direct experimental captures using (f1) $5\times$, (f3) $10\times$, and (f4) $20\times$ standard objectives, contrasted with (f2) the image recovered by the proposed RSCF. 
    (g1)-(g4) Corresponding processed and reconstructed field benchmarks for ROI (g) under identical modality configurations.}
    \label{figure 2}
\end{figure*}

To evaluate the performance of the proposed RSCF, a custom transmission-mode FPM system was conducted \cite{jiang2023spatial}. Figure \ref{figure 2}(b) illustrates the physical experimental setup, with its corresponding three-dimensional schematic detailed in Fig. \ref{figure 2}(a). The illumination module consists of a programmable $32 \times 32$ surface-mounted RGB LED array (Adafruit, 607), featuring 1024 distinct sources spaced 4 mm apart. The optical imaging train is composed of a low-magnification objective lens (Olympus LMPlanFLN $5\times$/0.13~NA) paired with a tube lens (Thorlabs AC254-150-C). An industrial CMOS Camera (Omron STC-MBS52POE, pixel size 3.45~$\mu$m) serves as the digital image acquisition sensor. To facilitate precision alignment, the LED array is integrated onto a multi-axis translation stage providing four degrees of freedom ($x,y,z$ translations and $z$-axis rotation). 

Synchronous coordination between the pattern illumination and sequential image acquisition is mediated by an Arduino Mega 2560 microcontroller. The microcontroller sequentially triggers the individual LED elements while controlling the camera exposure via external transistor-transistor logic (TTL) signals. Although this hardware-triggered synchronization scheme introduces a minor latency compared to free-running models, it permits adaptive exposure settings across different illumination angles. This capability ensures an optimal signal-to-noise ratio (SNR) by extending the integration time for weak, high-frequency dark-field measurements while preventing saturation in bright-field channels. The iterative reconstruction algorithm was implemented in MATLAB (R2022a) and executed on a portable computational workstation equipped with an NVIDIA GeForce RTX 4060 GPU (32~GB RAM) to accelerate the parallel matrix operations and the singular value decomposition loops.

\subsection{Full-FOV FPM Imaging}
The full-field-of-view (full-FOV) reconstruction capability of the proposed RSCF was validated by imaging a commercially prepared plant stem slice (Fig. \ref{figure 2}(c)). The programmable LED array was placed approximately 92 mm directly beneath the sample plane, and red illumination with a wavelength of 632~nm was employed. The light transmitted through the specimen was collected by the objective lens and relayed to the camera sensor through a tube lens. The adaptive exposure times were set to 200~ms for bright-field illumination and extended to 500~ms for dark-field modes in this part. A total of 225 low-resolution intensity images (each spanning 624 $\times$ 624 pixels) were acquired, corresponding to an effective field of view of 1.3 $\times$ 1.3~mm$^2$. The raw captured sequence is compiled in Fig. \ref{figure 2}(d), where the red boundary demarcates the theoretical bright-field region. Notably, a pronounced vignetting effect is visible near the bright-field boundary, manifesting as an uncalibrated spatial intensity decay.

Figure \ref{figure 2}(e) shows a representative low-resolution raw image under the central illumination alongside the full-FOV reconstruction using the proposed RSCF. For this implementation, the subspace decomposition operator $\Psi$ was instantiated via SVD to construct the orthogonal projection matrix $\mathcal{P}$. The detailed mathematical derivations of the residual-matrix principal component filtering and its associated gradient backpropagation steps are provided in Sections S1 and S3 of the Supplemental Document.

To provide a rigorous resolution comparison, two distinct regions of interest (ROIs), labeled (f) and (g) within the vascular bundle structures, are magnified in the lower panels of Fig. \ref{figure 2}. The RSCF reconstruction results [Figs. \ref{figure 2}(f2) and (g2)] are evaluated against direct images captured under conventional bright-field objectives of increasing power: a $5\times$/0.13~NA objective [Figs. \ref{figure 2}(f1) and (g1)], a $10\times$/0.25~NA objective [Figs. \ref{figure 2}(f3) and (g3)], and a $20\times$/0.40~NA objective [Figs. \ref{figure 2}(f4) and (g4)]. While conventional low-magnification objectives provide a wide FOV at the expense of structural resolution, high-magnification objectives resolve fine details but suffer from a severely restricted imaging throughput. In contrast, the proposed RSCF simultaneously yields a high spatial resolution comparable to a high-NA lens while maintaining the wide FOV inherent to the $5\times$ objective. Furthermore, as highlighted in Figs. \ref{figure 2}(f3) and (f4), conventional high-NA imaging suffers from localized defocusing artifacts due to slight axial thickness variations across the sample slice. The RSCF reconstruction demonstrates an inherently superior depth-of-field tolerance to minor sample thickness fluctuations, preserving high contrast across the entire field without localized blurring.
 
In this experimental geometry, the synthetic NA is governed by the sum of the objective lens NA and the maximum illumination angle subtended by the active LED elements. Backed by the $5\times$/0.13~NA objective, the proposed RSCF achieves an equivalent synthetic aperture of approximately $\text{NA}_{\text{sys}} \approx 0.42$, representing a $3.24\times$ enhancement over the physical diffraction limit. Crucially, the experimental results demonstrate that fine sub-diffraction details are successfully recovered without applying any raw image preprocessing to correct for the severe vignetting. Because systematic anomalies like vignetting, LED position shifts, and system aberrations exhibit strong spatial coherence, they are natively isolated within the dominant singular modes of the residual matrix and neutralized by the subspace projection matrix $\mathcal{P}$. Consequently, the proposed optimization scheme exhibits exceptional robustness against coupled environmental perturbations, matching the fidelity of high-power objectives while providing a robust, calibration-relaxed alternative for practical, large-scale computational optical microscopy.

\section{Discussion}
\subsection{Evaluation Under Severe Coupled Model Mismatches}

Conventional FPM reconstruction algorithms are inherently constrained by their reliance on the accuracy of the forward physical model. Although advanced correction schemes, such as EPRY algorithm \cite{Ou:14}, regularized loss functions \cite{Shi:21}, and simulated annealing methods \cite{Sun:16}, have mitigated isolated systematic errors, they often fail when confronted with highly coupled, mixed degradations encountered in real-world environments. Feature-domain optimization, represented by FD-FPM \cite{Zhang:24}, offers an alternative by evaluating errors within transformed spaces to suppress multiple error mechanisms simultaneously. However, these frameworks lack a selective filtering mechanism to distinguish physically consistent forward deviations from random high-frequency fluctuations, leaving room for further optimization under low signal-to-noise ratio (SNR) conditions.

To overcome these limitations, our framework exploits the distinct low-rank and sparse properties exhibited by systematic forward-model mismatches within the residual subspace. Unlike stochastic noise, which distributes evenly across the singular value spectrum, systematic deviations (e.g., vignetting, aberrations, and source misalignments) manifest as highly correlated spatial structures that concentrate within dominant singular modes.

\begin{figure}[t!]
    \centering\includegraphics[width=\linewidth]{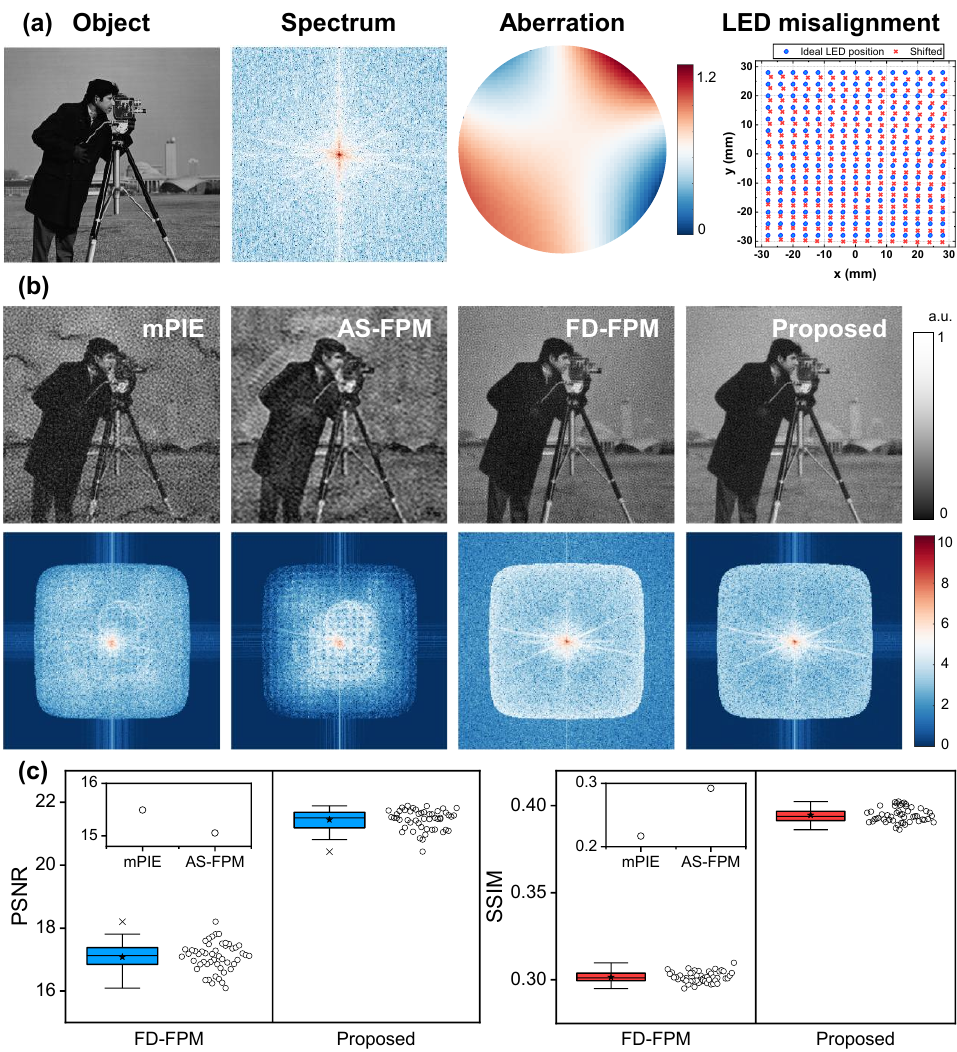}
    \caption{Reconstruction and evaluation of the low-quality simulation data under model mismatch. 
    (a) The complex amplitude object, Fourier spectrum, and the optical aberrations used to generate the low-quality data, as well as the position offsets of LED light in the $x-y$ direction. 
    (b) Comparison of the reconstructed images and Fourier spectrum using mPIE, AS-FPM, FD-FPM, and the proposed method.
    (c) Statistics of PSNR and SSIM scores for different reconstruction methods in 50 repeated simulation experiments.
    }
    \label{simulation}
\end{figure}
To evaluate the robustness of the proposed RSCF against such coupled perturbations, we generated a low-quality simulated dataset incorporating simultaneous random coefficient Zernike-fitted aberrations, overall and random LED misalignments, as well as additive Gaussian noise, as shown in Fig. \ref{simulation}(a). All stochastic components, denoted as randn(·) in Matlab, were generated using a fixed random seed to ensure strict reproducibility and the intensity is controlled by given coefficients. High-resolution complex reconstructions were performed using momentum-accelerated PIE (mPIE) \cite{Maiden:17}, adaptive step-size FPM (AS-FPM) \cite{Zuo:16}, FD-FPM \cite{Zhang:24}, and the proposed RSCF, with the results summarized in Fig. \ref{simulation}(b).  

It is found that conventional algorithms like mPIE and AS-FPM fail to converge under heavy coupled uncertainties, resulting in severe phase artifacts and noise amplification. While FD-FPM partially resolves the object topology, it remains susceptible to noise-induced background fluctuations. In contrast, the proposed RSCF recovers fine structural details with high contrast. This performance is supported by the combined scaling of residual-subspace projection a stochastic mini-batch optimization strategy, where introducing a random measurement subset during each iterative update helps smooth the non-convex error landscape \cite{masters2018revisitingsmallbatchtraining}. For a rigorous evaluation, 50 independent trial simulations were executed and the reconstruction quality was quantified via the Peak Signal-to-Noise Ratio (PSNR) and the Structural Similarity Index Measure (SSIM). The comparative analysis in Fig. \ref{simulation}(c) reveals that the proposed RSCF consistently outperforms alternative algorithms, maintaining a narrow, high-valued metric distribution across all trials. 

To uncover the underlying physical mechanism of the proposed RSCF, we tracked the dynamic evolution of the singular value spectrum during the iterative process. A primary characteristic of this residual subspace analysis lies in the mathematical entanglement within the dominant components: both the object reconstruction errors (inherent to the early phases of iterative optimization) and the physical systematic errors manifest as primary singular values within the residual domain. At the current stage, separating these two distinct error sources directly from a single subspace projection remains challenging. Consequently, the framework relies on the progressive convergence of the forward model to iteratively fit and decouple the object spectrum. As the object estimate refines, its corresponding fitting error in the residual domain gradually vanishes; the remaining persistent features left in the residual subspace are then directly attributed to the systematic errors of the imaging system. This decoupled transition from recovering coarse global topography to isolating system defects is verified and illustrated in Section S4A and Fig. S3 of the Supplemental Document, using a complex target composed of standard reference images.

Additional comparative results for diverse simulated samples are provided in Section S4B and Figs. S4–S6 of the Supplemental Document. Furthermore, because our method embeds pupil function retrieval directly into the iterative loop, it maintains robust performance even under large optical aberrations, as demonstrated in Section S4C and Figs. S7–S8 of the Supplemental Document. These collective simulation results show that RSCF provides superior reconstruction quality and numerical stability when processing highly degraded data, significantly outperforming existing baseline methods.

\subsection{Low-Rank Subspace Verification}
The performance of the selection operator $\mathcal{S}$ is governed by the truncation threshold $\sigma$, which balances structural error representation against random noise suppression within the residual matrix. Mapping this threshold $\sigma$ to the cumulative variance contribution rate (CVCR) determines the number of retained principal components, $k$, during localized subspace projection. Identifying the optimal singular-spectrum cutoff represents a classic trade-off in low-rank matrix estimation \cite{jolliffe2016principal}; Retaining too few modes discards physical high-frequency updates, whereas over-retaining components allows stochastic perturbations to destabilize the optimization path.
\begin{figure}[t!]
    \centering\includegraphics[width=\linewidth]{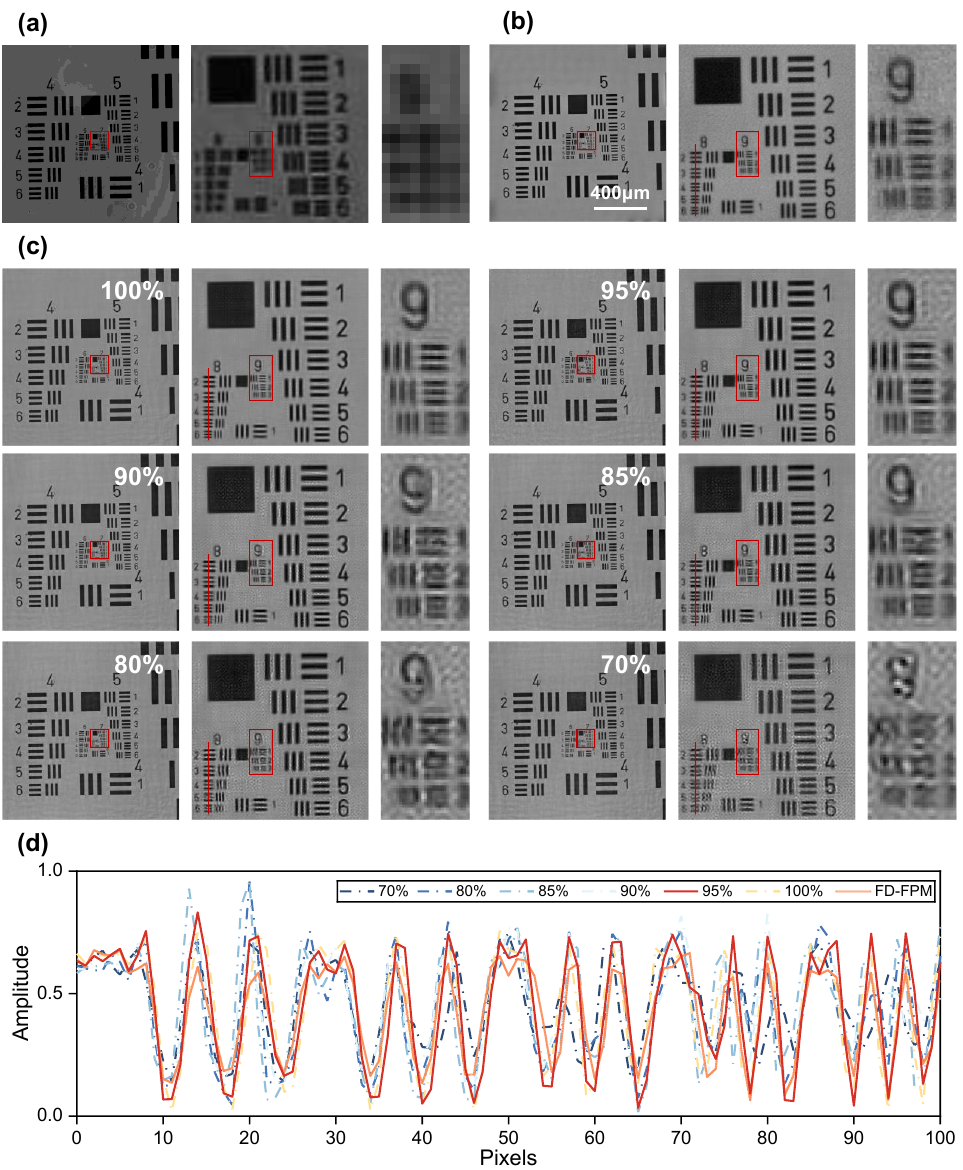}
    \caption{Comparison of reconstruction results under different principal component contents determined by the cumulative variance contribution rate.
    (a) The original image and the selected and magnified sub-regions; (b) The reconstruction result by FD-FPM; (c) The reconstruction results in RSCF with $k$ determined by different CVCRs; (d) Comparison diagram of the linear profile curves of Figure (b)-(c).
    }
    \label{figure 5}
\end{figure}

To systematically evaluate this dependency, we performed high-resolution reconstructions across a CVCR range from $70\%$ to $100\%$, as summarized in Fig. \ref{figure 5}. When the CVCR is set below $80\%$, the restricted rank of the subspace prevents the framework from capturing high-order systematic variations, leaving the Group 9 resolution elements unresolved [Fig. \ref{figure 5}(c)]. Increasing the CVCR to $90\%$ recovers basic structural features. At an optimal CVCR of $95\%$, the framework achieves peak fidelity: the Group 9 elements are clearly recovered, and the line profile across Group 8 [Fig. \ref{figure 5}(d)] demonstrates optimal contrast. 

Crucially, this $95\%$ threshold aligns with the physical separation mechanism discussed in the previous section: it provides sufficient rank to capture both the progressively updated object spectrum and the persistent, low-rank systematic errors, while effectively cutting off the remaining $5\%$ of energy dominated by isotropic noise. Conversely, when the CVCR reaches $100\%$, the selection operator retains the entire singular spectrum ($k = \text{batch size}$), causing subspace projection to degenerate into a standard spatial Euclidean metric. At this limit, the unconstrained accumulation of random noise and coupled experimental artifacts severely degrades reconstruction quality, inducing low-frequency dark-band distortions near the FOV boundaries.

The empirical relationship between the target CVCR and the dynamically assigned rank $k$ within a sub-batch size of 45 is detailed in Table \ref{tab:my_label}. The statistical distribution reveals that a compact subset of principal components ($\text{mean } k = 13.15 \pm 5.26$ for a $95\%$ CVCR) captures the vast majority of variance within the residual matrix. This energy concentration confirms that the structured forward-model mismatches occupy a highly compressed subspace, validating the core physical assumption of the residual-subspace filtering strategy.

\begin{table*}
    \centering
    \begin{tabular}{ccccccc}
        \hline
           CVCR&  70\%&  80\%&  85\%&  90\%&  95\%& 100\%\\
       \hline
           $k$ (Mean$\pm$Std)&  5.18$\pm$1.77&  6.78$\pm$2.48&  8.13$\pm$3.14&  9.79$\pm$3.85&  13.15$\pm$5.26& 45\\
           $k$ (Median)&  5&  7&  8&  9&  12& 45\\
       \hline
    \end{tabular}
    \caption{The relationship between CVCR and the $k$ value in the small batch size of 45.}
    \label{tab:my_label}
\end{table*}

\subsection{Experimental Resolution and Adaptability}

\begin{figure*}[htb!]
\centering\includegraphics[width=\linewidth]{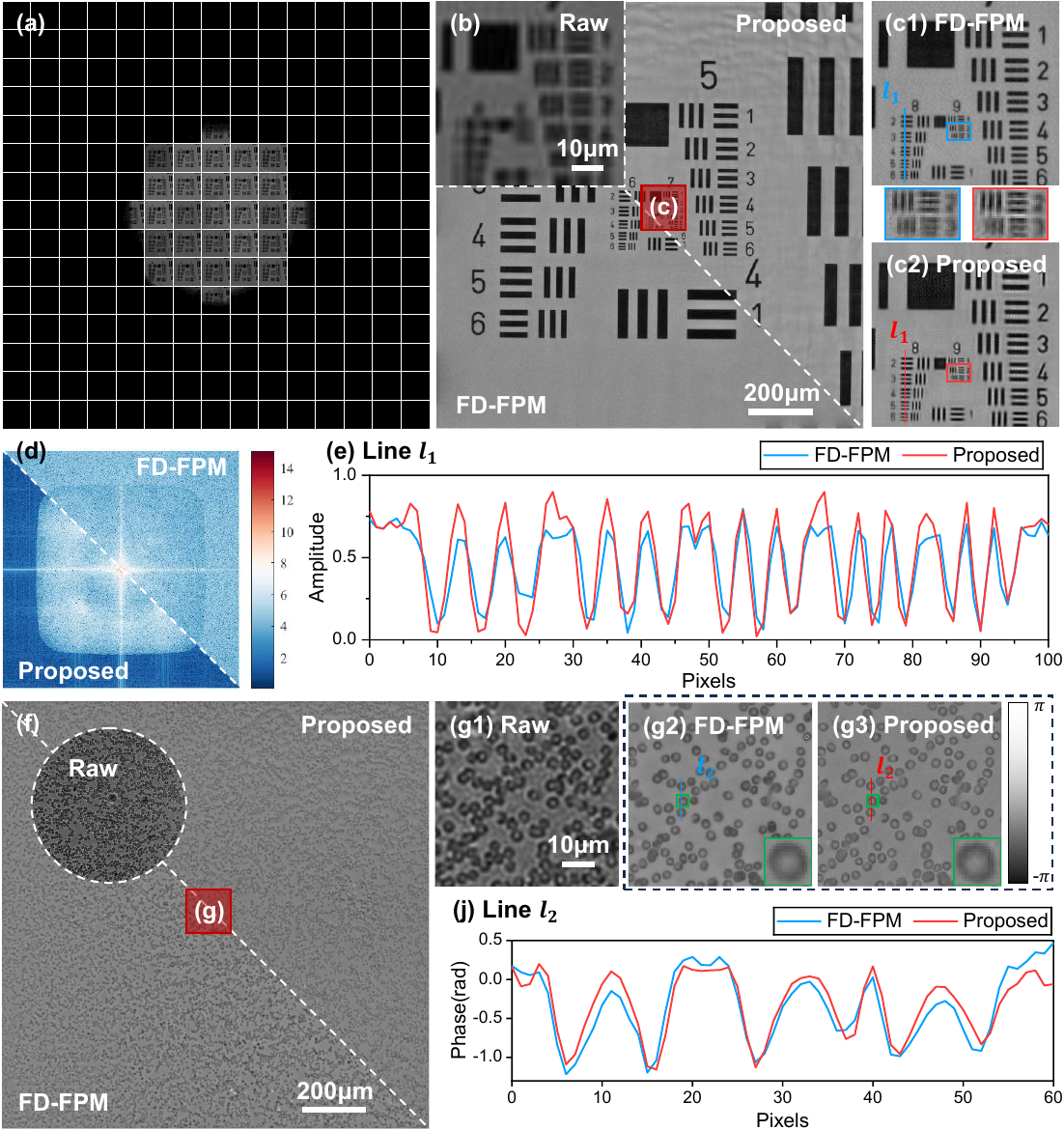}
    \caption{Full-FOV reconstruction of the USAF1951 resolution target and blood smear samples under $15\times15$ LED red illumination. 
    (a) Acquisition of a series of low-resolution images affected by the vignetting effect. 
    (b) Comparison of the results of full-FOV reconstruction, obtained using the FD-FPM and RSCF. The upper left part shows the region (c) slice of the low-resolution image obtained by the central LED; 
    (c1)-(c2) are enlarged images of sub-region (c), including the theoretical resolution of the system low-resolution image and the theoretical resolution of FPM reconstruction. 
    (d) Comparison of the Fourier spectrum images obtained by the two reconstruction methods. 
    (e) Amplitude curve analysis along the line $l_1$ of the reconstructed images (c1)-(c2). 
    (f) Comparison of the low-resolution image obtained by the system center LED for the bloodsmear and the results of full-FOV reconstruction using FD-FPM and RSCF. 
    (g1)-(g3) are enlarged images of region (g), where (g1) is the amplitude image directly obtained, and (g2)-(g3) are the reconstructed phase images. 
    (j) Phase curve analysis along the line $l_2$ of the reconstructed images (g2)-(g3).
    }
    \label{resolution}
\end{figure*}

The experimental resolution limits and cross-modality adaptability of the custom-built FPM platform were evaluated by imaging a standard USAF1951 resolution target. As shown in the raw, low-resolution central-illumination frame [Fig. \ref{resolution}(b)], the physical diffraction limit of the system restricts the observable elements to Group 7, elements 5 ($2.43$ $\mu\text{m}$). Figures \ref{resolution}(c1) and \ref{resolution}(c2) display the magnified full-FOV amplitude reconstructions achieved by FD-FPM and the proposed RSCF, respectively. Both methods demonstrate substantial synthetic aperture expansion, clearly resolving the fine lines of Group 9, element 2 ($0.87$ $\mu\text{m}$), with Group 9, element 3 approaching the theoretical cut-off frequency of the synthesized aperture ($\text{NA}_{\text{sys}} \approx 0.42$).

Crucially, the line-profile analysis along $l_1$ in Fig. \ref{resolution}(e) reveals that the proposed RSCF yields a markedly higher modulation contrast than FD-FPM. This performance is a direct consequence of the subspace parameter established in our previous analysis: by implementing the optical $95\%$ CVCR cutoff, the proposed framework retains sufficient singular modes to reconstruct high-frequency specimen details while blocking the remaining $5\%$ of energy dominated by random measurement noise. Consequently, the proposed RSCF exhibits exceptional resilience against the heavy, space-invariant vignetting visible near the FOV boundary [Fig. \ref{resolution}(a)] without requiring independent flat-field pre-calibration.

To test the framework beyond pure amplitude modulation, we evaluated its practical utility on complex phase specimens, which present severe scattering and diffraction challenges in label-free biomedical imaging.  Figure \ref{resolution}(f) shows the central-illumination image of a dense blood smear along with its reconstructed phase maps. Even in regions obscured by cellular clustering and out-of-focus background impurities, the characteristic biconcave morphology of individual erythrocytes is clearly resolved in the RSCF phase distributions [Figs. \ref{resolution}(g3)].

This experimental robustness underscores the physical decoupling mechanism detailed in our simulation section. During the iterative process, the high-order phase distortions from out-of-focus impurities and systematic vignetting naturally cluster as low-rank components within the dominant residual subspace. As the forward model progressively fits and decouples the true cell spectrum, these persistent environmental aberrations are isolated, preventing them from bleeding into the image updates. As a result, the quantitative phase profile across line $l_2$ [Fig. 4(j)] demonstrates that our method effectively eliminates phase-wrapping artifacts and low-frequency background fluctuations, confirming its high fidelity for complex biological imaging under non-ideal experimental environments.

\section{Conclusion}
In summary, we developed the RSCF, a robust and calibration-relaxed reconstruction paradigm for COI, and validated its performance on an FPM platform. By shifting the optimization metric from the conventional pixel-wise spatial domain to a low-rank feature subspace, the RSCF mitigates the core limitations of the conventional $L_2$-norm minimization. This approach allows the optimization loop to separate and prioritize deterministic, low-rank error components over isotropic, random fluctuations. Comprehensive numerical and experimental validations against state-of-the-art methods, such as FD-FPM, demonstrate that RSCF provides high reconstruction fidelity and convergence stability under severe, coupled system perturbations without requiring hardware modifications or analytical pre-calibration.

The efficacy of the proposed RSCF is rooted in its physically informed exploitation of optimization residuals. Unlike standard metrics that minimize total residual energy indiscriminately, RSCF recognizes that systematic forward-model discrepancies, including vignetting, optical aberrations, LED intensity fluctuations, and LED positional misalignments, possess strong spatio-spectral correlations.  Consequently, these systematic errors naturally cluster within a compact set of dominant singular vectors. By executing gradient updates within this low-rank subspace, the algorithm selectively isolates and rectifies structural modeling flaws while suppressing noise amplification. This selective mechanism guarantees physically plausible phase and amplitude updates for both test targets and complex biological specimens.

Nevertheless, certain constraints of the proposed RSCF require further development. A primary limitation lies in the mathematical entanglement within the dominant subspace components: both the object reconstruction errors (inherent to the iterative optimization process) and the physical systematic errors manifest as primary singular values within the residual domain. Currently, directly decoupling these two distinct error sources within the subspace projection remains difficult; the framework thus relies on the progressive convergence of the forward model to iteratively resolve the object spectrum, leaving the persistent residual features to be attributed to systematic discrepancies. Furthermore, performing iterative SVD within a stochastic gradient descent loop introduces non-trivial computational overhead that scales with the matrix dimensions. The framework also relies on an empirical selection threshold for the singular value cutoff, which under extremely low signal-to-noise conditions may become ambiguous as the noise floor elevates and overlaps with trailing systematic error modes.

Future efforts will address these challenges by exploring randomized, low-complexity matrix decomposition algorithms to accelerate the computing loop, alongside developing information-theoretic models for automated threshold determination. Beyond its application to FPM, the core philosophy of subspace-driven residual analysis is highly modular. The strategy of leveraging latent residual structures to bridge the gap between forward models and physical reality can be readily extended to other computational optical modalities, including ptychographic tomography, structured illumination microscopy, and single-photon imaging. We anticipate that this residual-centric optimization paradigm will provide a robust tool for solving complex inverse problems in non-ideal experimental environments.

\begin{backmatter}
\bmsection{Funding} Basic and Applied Basic Research Foundation of Guangdong
Province (2026A1515011642, 2023B1515040023); Guangzhou Science and Technology Program
(202201011671).

\bmsection{Disclosures} The authors declare no conflicts of interest.

\bmsection{Data availability} Data underlying the results presented in this paper are not publicly available at this time but may be obtained from the authors upon reasonable request.

\bmsection{Supplemental Document} See Supplement 1 for supporting content.

\end{backmatter}



\bibliography{Reference}

\end{document}